\documentclass[graybox]{svmult}

\pdfoutput=1

\usepackage{mathptmx}       % selects Times Roman as basic font
\usepackage{helvet}         % selects Helvetica as sans-serif font
\usepackage{courier}        % selects Courier as typewriter font
\usepackage{type1cm}        % activate if the above 3 fonts are
\usepackage{makeidx}         % allows index generation
\usepackage{graphicx}        % standard LaTeX graphics tool
\usepackage{multicol}        % used for the two-column index
\usepackage[bottom]{footmisc}% places footnotes at page bottom

\usepackage{enumitem}
\usepackage{amssymb}

\makeindex             % used for the subject index
\begin{document}

\title*{Misinformation spreading on Facebook}
\author{Fabiana Zollo and Walter Quattrociocchi}
\institute{Fabiana Zollo \at Ca' Foscari University of Venice, Via Torino, 155 - 30172, Venice, Italy\\ \email{fabiana.zollo@unive.it}
\and Walter Quattrociocchi \at IMT School for Advanced Studies, Piazza San Francesco, 19 - 55100 Lucca, Italy\\ \email{walter.quattrociocchi@imtlucca.it}}
\maketitle

%SPREADING DYNAMICS IN SOCIAL SYSTEMS

\abstract{Social media are pervaded by unsubstantiated or untruthful rumors, that contribute to the alarming phenomenon of misinformation. The  widespread presence of a heterogeneous mass of information sources may affect the mechanisms behind the formation of public opinion. Such a scenario is a florid environment for digital wildfires when combined with functional illiteracy, information overload, and confirmation bias. In this essay, we focus on a collection of works aiming at providing quantitative evidence about the cognitive determinants behind misinformation and rumor spreading. We account for users' behavior with respect to two distinct narratives: \textit{a)} conspiracy and \textit{b)} scientific information sources. In particular, we analyze Facebook data on a time span of five years in both the Italian and the US context, and measure users' response to \textit{i)} information consistent with one's narrative, \textit{ii)} troll contents, and  \textit{iii)} dissenting information e.g., debunking attempts.
Our findings suggest that users tend to a) join polarized communities sharing a common narrative (\textit{echo chambers}), b) acquire information confirming their beliefs (\textit{confirmation bias}) even if containing false claims, and c) ignore dissenting information.}

\section{Introduction}
\label{sec:intro}

The rapid advance of the Internet and web technologies facilitated global communications all over the world, allowing news and information to spread rapidly and intensively. These changes led up to the formation of a new scenario, where people actively participate in both contents' production and diffusion, without the mediation of journalists or experts in the field. The emergence of such a wide, heterogeneous (and disintermediated) mass of information sources may affect contents' quality and the mechanisms behind the formation of public opinion \cite{lazarsfeld1968, katz1970, watts2007}. Indeed, despite the enthusiastic rhetoric about collective intelligence \cite{levy1999},  unsubstantiated or untruthful rumors reverberate on social media, contributing to the alarming phenomenon of misinformation.  
Since 2013, the World Economic Forum (WEF) has been placing the global danger of massive digital misinformation at the core of other technological and geopolitical risks, ranging from terrorism, to cyber attacks, up to the failure of global governance \cite{howell2013}. People are misinformed when they hold beliefs neglecting factual evidence, and misinformation may influence public opinion negatively. Empirical investigations have showed that, in general, people tend to resist facts, holding inaccurate factual beliefs confidently \cite{kuklinski2000misinformation}. Moreover, corrections frequently fail to reduce misperceptions \cite{nyhan2010} and often act as a \textit{backfire effect}.

Thus, beyond its undoubted benefits, a hyperconnected world may allow the viral spread of misleading information, which may have serious real-word consequences. In that direction, examples are numerous. Inadequate health policies in South Africa led to more than 300,000 unnecessary AIDS deaths \cite{moore2009}, however the events were exacerbated by AIDS \textit{denialists}, who state that HIV is inoffensive and that antiretroviral drugs cause, rather than treat, AIDS. Similar considerations may be extended to the  Ebola outbreak in west Africa: after the death of two people having drunk salt water, the World Health Organisation (WHO) had to restate that all rumours about hypothetical cures or practices were false and that their use could be dangerous \cite{WHO2014}. Or again, the American case of Jade Helm 15, a military training exercise which took place in multiple US states, but turned out to be perceived as a conspiracy plot aiming at imposing martial law, to the extent that Texas Gov. Greg Abbott ordered the State Guard to monitor the operations. 

Certainly, such a scenario represents a florid environment for digital wildfires, especially when combined with functional illiteracy, information overload, and \textit{confirmation bias} -- i.e., the tendency to seek, select, and interpret information coherently with one's system of beliefs \cite{nickerson1998confirmation}. On the Internet people can access always more extreme versions of their own opinions. In this way, the benefits coming from the exposure to different points of views can be dramatically reduced \cite{lessig2009code}. Individuals, and the groups that they form, may move to a more extreme point in the same direction indicated by their own preexisting beliefs; indeed, when people discuss with many like-minded others, their views become more extreme \cite{sunstein2002law}. First evidences of social contagion and misperception induced by social groups emerged in the famous experiment conducted by Solomon Asch in 1955 \cite{asch1955opinions}. The task of the participants was very simple: they had to match a certain line placed on a white card with the corresponding one (i.e., having the same length) among three other lines placed on another white card. The subject was one of the eight people taking part to the test, but was unaware that the others were there as part of the research. The experiment consisted of three different rounds. In the first two rounds everyone provided the right (and quite obvious) answer. In the third round some group members matched the reference line to the shorter or longer one on the second card, introducing the so-called \textit{unexpected disturbance} \cite{kahan1997social}. Normally subjects erred less than 1\% of the time; but in the third case they erred 36.8\% of the time \cite{akerlof1996analysis}. Another relevant study was conducted by James Stoner, who identified the so-called \textit{risky shift} \cite{stoner1968risky}. In the experiment people were first asked to study twelve different problems and provide their solution; after that, they had to take a final decision together, as a group. Out of thirteen groups, twelve repeatedly showed a pattern towards greater risk-taking.

Misinformation, as well of rumor spreading, deals with these and several other aspects of social dynamics. However, adoption and contagion are often illustrated under the oversimplified metaphor of the \textit{virus}: ideas spread by ``contact" and people ``infected" become active spreaders in the contagion process. We believe that such a metaphor is misleading, unless we consider that the receptor of such a virus is complex and articulated. Indeed, the adoption of ideas and behaviors deals with a multitude of cognitive dimensions, such as intentionality, trust, social norms, and confirmation bias. Hence, simplistic models adapted from mathematical epidemiology are not enough to understand social contagion. It is crucial to focus on such relevant research questions by using methods and applying tools that go beyond the pure, descriptive statistics of big data. In our view, such a challenge can be addressed by implementing a cross-methodological, interdisciplinary approach which takes advantage of both the question-framing capabilities of social sciences and the experimental and quantitative tools of hard sciences.

\section{Outline}
The chapter is structured as follows. In section \ref{sec:tools} we provide the background of our research work, as well as tools and methodology adopted; in section \ref{sec:datasets} we describe the datasets; in section \ref{sec:echo} we discuss the dynamics behind information consumption and the existence of echo chambers on both the Italian and the US Facebook; in section \ref{sec:cascades} we show how confirmation bias dominates information spreading; in section \ref{sec:troll} we focus on users' interaction with paradoxical and satyrical information (trolls), while in section \ref{sec:debunking} we analyze users' response to debunking attempts. In section \ref{sec:emotional} we target the emotional dynamics inside and across echo chambers. Finally, we draw our conclusions in section \ref{sec:conclusions}.

\section{Background and Research Methodology}
\label{sec:tools}
In 2009 a paper on Science \cite{lazer2009life} proclaims the birth of the Computational Social Science (CSS), an emerging research field aiming at studying massive social phenomena quantitatively, by means of a multidisciplinary approach based on Computer Science, Statistics, and Social Sciences. Since CSS benefits from the large availability of data from online social networks, it is attracting researchers in ever-increasing numbers as it allows for the study of mass social dynamics at an unprecedented level of resolution. Recent studies have pointed out several important results ranging from social contagion \cite{ugander2012, mcpherson2001, aral2009} up to information diffusion \cite{adar2004implicit, bakshy2011everyone}, passing through the virality of false claims \cite{cheng2014,dow2013}. A wide literature branch is also devoted to understanding the spread of rumors and behaviors by focusing on structural properties of social networks to determine the way in which news spread in social networks, what makes messages go viral, and what are the characteristics of users who help spread such information \cite{cheng2014,dow2013,centola2010,ugander2012}. Several works investigated how social media can shape and influence the public sphere \cite{adamic2005political,bakshy2015exposure,conover2011political,conover2011predicting}, and efforts to contrast misinformation spreading range from algorithmic-based solutions up to tailored communication strategies \cite{almansour2014model,ciampaglia2015computational,gupta2014tweetcred,qazvinian2011rumor,ratkiewicz2011detecting,resnick2014rumorlens}.

Along this path, important issues have been raised around the emergence of the \textit{echo chambers}, enclosed systems where users are exposed only to information coherent with their own system of beliefs \cite{sunstein2009republic}. Many argue that such a phenomenon is directly related to the algorithms used to rank contents \cite{pariser2011filter}. Speaking of this, Facebook research scientists quantified exactly how much individuals can be exposed to ideologically diverse news and information on social media \cite{bakshy2015exposure}, finding that individual's choice about contents has an effect stronger than that of Facebook's News Feed algorithm in limiting the exposure to cross-cutting content. Undoubtedly, selective exposure to specific contents facilitates the aggregation of users in echo chambers, wherein external and contradicting versions are ignored \cite{knobloch2012selective}. Moreover, the lack of experts mediating the production and diffusion of content may encourage speculations, rumors, and mistrust, especially on complex issues. Pages about conspiracy theories, chem-trails, reptilians, or the link between vaccines and autism, proliferate on social networks, promoting alternative narratives often in contrast to mainstream content. Thus, misinformation online is pervasive and difficult to correct. To face the issue, several algorithmic-driven solutions have been proposed both by Google and Facebook \cite{dong2015knowledge,fb2015hoaxes}, that joined other major corporations to provide solutions to the problem and try to guide users through the digital information ecosystem \cite{firstdraftnews}. Simultaneously, it has also been observed the rapid spread of blogs and pages devoted to debunk false claims, namely \textit{debunkers}. 

Moreover, the diffusion of unreliable content may lead to confuse unverified stories with their satirical counterparts. Indeed, it has been noticed the proliferation of satirical, wacky imitations of conspiracy theses. In this regard, there is a large community of people, known as \textit{trolls}, behind the creation of Facebook pages aimed at diffusing caricatural and paradoxical contents mimicking conspiracy news. Their activities range from controversial comments and satirical posts, to the fabrication of purely fictitious statements, heavily unrealistic and sarcastic. According to Poe's law \cite{aikin2013poe}, without a blatant display of humor, it is impossible to create a parody of extremism or fundamentalism that someone won't mistake for the real thing. Hence, trolls are often accepted as realistic sources of information and, sometimes, their memes become viral and are used as evidence in online debates from real political activists. As an example, we report one of the most popular memes in Italy: 
\begin{quotation}
Italian Senate voted and accepted (257 in favor, 165 abstained) a law proposed by Senator Cirenga aimed at providing politicians with a ¤134 Billion fund to help them find a job in case of defeat in the next political competition.
\end{quotation}
It would be easy to verify that the text contains at least three false statements: \textit{i)} Senator Cirenga does not exist and has never been elected in the Italian Parliament, \textit{ii)} the total number of votes is higher than the maximum possible number of voters, and \textit{iii)} the amount of the fund corresponds to more than 10\% of Italian GDP. Indeed, the bill is false and such a meme was created by a troll page. Nonetheless, on the wave of public discontent against Italian policy-makers, it quickly became viral, obtaining about 35K shares in less than one month. Nowadays, it is still one of the most popular arguments used by protesters manifesting all over Italian cities.

Such a scenario makes crucial the quantitative understanding of the social determinants related to content selection, news consumption, and beliefs formation and revision. In this essay, we focus on a collection of works \cite{bessi2015science,bessi2015viral,bessi2016homophily,del2016spreading,zollo2015debunking,zollo2015emotional} aiming at characterizing the role of confirmation bias in viral processes online. We want to investigate the cognitive determinants behind misinformation and rumor spreading by accounting for users' behavior on different and specific narratives. In particular, we define the domain of our analysis by identifying two well-distinct narratives: \textit{a)}conspiracy and \textit{b)} scientific information sources. Notice that we do not focus on the quality or the truth value of information, but rather on its verifiability. While producers of scientific information as well as data, methods, and outcomes are readily identifiable and available, the origins of conspiracy theories are often unknown and their content is strongly disengaged from mainstream society and sharply divergent from recommended practices. 

Thus, we first analyze users' interaction with Facebook pages belonging to such distinct narratives on a time span of five years (2010-2014), in both the Italian and the US context. Then, we measure users' response to \textit{i)} information consistent with one's narrative,  \textit{ii)} troll contents, and  \textit{iii)} dissenting information e.g., debunking attempts. 

\section{Datasets}
\label{sec:datasets}

We identify two main categories of pages: conspiracy news -- i.e. pages promoting contents \textit{neglected} by mainstream media -- and science news. The first category includes all pages diffusing conspiracy information (i.e., pages that disseminate controversial information, most often lacking supporting evidence and sometimes contradictory of the official news). Pages like {\em I don't trust the government}, {\em Awakening America}, or {\em Awakened Citizen} promote heterogeneous contents ranging from aliens, chem-trails, geocentrism, up to the causal relation between vaccinations and homosexuality. The second category is that of scientific dissemination and includes institutions, organizations, scientific press having the main mission to diffuse scientific knowledge. For example, pages like {\em Science}, {\em Science Daily}, and {\em Nature} are active in diffusing posts about the most recent scientific advances. Finally, we identify two additional categories of pages: 
\begin{enumerate}
\item Troll: sarcastic, paradoxical messages mocking conspiracy thinking (for the Italian dataset);
\item Debunking: information aiming at correcting false conspiracy theories and untruthful rumors circulating online (for the US dataset).
\end{enumerate}
To produce our datasets, we built a large atlas of Facebook public pages with the assistance of several groups (\textit{Skepti Forum}, \textit{Skeptical spectacles}, \textit{Butac}, \textit{Protesi di Complotto}), which helped in labelling and sorting both conspiracy and scientific sources. To validate the list, all pages have then been manually checked by looking at their self-description and the type of promoted content. The exact breakdowns of the Italian and US Facebook datasets are reported in Table~\ref{tab:dataIT} and Table~\ref{tab:dataUS}, respectively. The entire data collection process is performed exclusively by means of the Facebook Graph API \cite{fb_graph_api}, which is publicly available and can be used through one's personal Facebook user account. We used only public available data (users with privacy restrictions are not included in our dataset). Data was downloaded from public Facebook pages that are public entities. Users' content contributing to such entities is also public unless users' privacy settings specify otherwise and in that case it is not available to us. When allowed by users' privacy specifications, we accessed public personal information. However, in our study we used fully anonymized and aggregated data. We abided by the terms, conditions, and privacy policies of Facebook.

\begin{table}
\caption{Breakdown of the Italian Facebook dataset.}
\label{tab:dataIT}
\begin{tabular}{p{3.4cm}p{2.6cm}p{2.6cm}p{2.6cm}}
\hline\noalign{\smallskip}
 & Science & Conspiracy & Troll  \\
\noalign{\smallskip}\svhline\noalign{\smallskip}
Pages & 34 & 39 & 2\\
Posts & 62,705 & 208,591 & 4,709\\
Likes & 2,505,399 & 6,659,382 & 40,341\\
Comments & 180,918 & 836,591 & 58,686\\
Likers & 332,357  & 864,047 & 15,209\\
Commenters & 53,438 & 226,534 & 43,102\\
\noalign{\smallskip}\hline\noalign{\smallskip}
\end{tabular}
\end{table}

\begin{table}
\caption{Breakdown of the US Facebook dataset.}
\label{tab:dataUS}
\begin{tabular}{p{3.4cm}p{2.6cm}p{2.6cm}p{2.6cm}}
\hline\noalign{\smallskip}
 & Science & Conspiracy & Debunking  \\
\noalign{\smallskip}\svhline\noalign{\smallskip}
Pages & 83 & 330 & 66\\
Posts & 262,815 & 369,420 & 47,780\\
Likes & 453,966,494 & 145,388,117 & 3,986,922\\
Comments & 22,093,692 & 8,304,644 & 429,204\\
Likers & 39,854,663 & 19,386,131 & 702,122\\
Commenters & 7,223,473 & 3,166,726 & 118,996\\
\noalign{\smallskip}\hline\noalign{\smallskip}
\end{tabular}
\end{table}

\section{Echo Chambers}
\label{sec:echo}

\subsection{Attention Patterns}
We start our discussion by analyzing how information gets consumed by users in both the Italian \cite{bessi2015science,bessi2015viral,bessi2016homophily} and the US Facebook \cite{zollo2015debunking}. As a first step, we focus on users' actions allowed by Facebook's interaction paradigm i.e., likes, comments, and shares. Each action has a particular meaning \cite{ellison2007benefits}: while a \textit{like} represents a positive feedback to the post, a \textit{share} expresses the desire to increase the visibility of a given information; finally, a \textit{comment} is the way in which the debate takes form around the topic of the post. Also, we consider the lifetime of a post (respectively, a user) i.e., the temporal distance between the first and last comment to the post (respectively, of the user). We also define the persistence of a post (respectively, a user) as the Kaplan-Meier estimates of survival functions by accounting for the lifetime of the post (respectively, the user). 

\begin{figure}
\centering
\includegraphics[width=.7\textwidth]{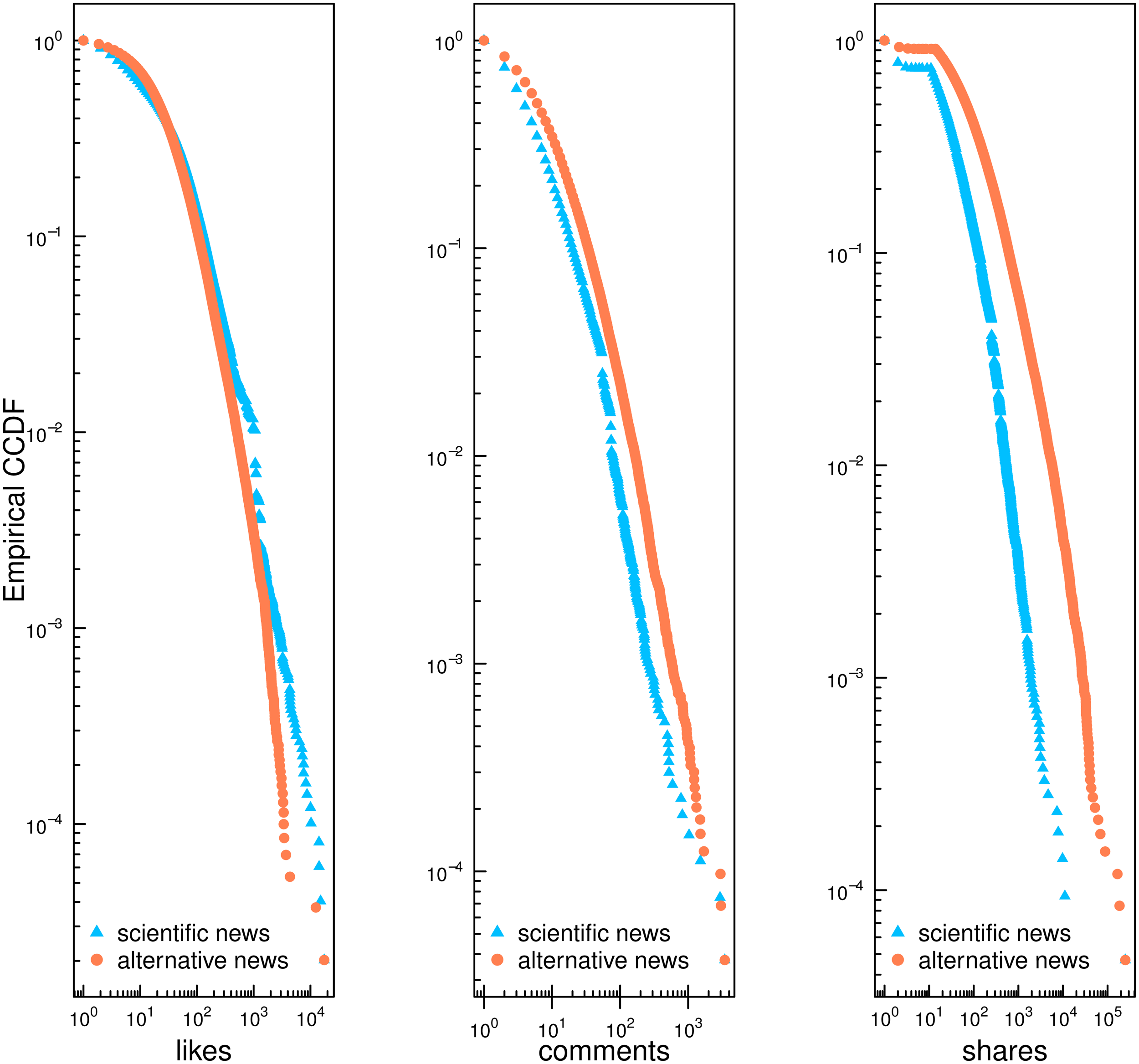}
\caption{\textsc{Italian Facebook}. Empirical complementary cumulative distribution functions (CCDFs) of users' activity (likes, comments and shares) on posts grouped by category. Distributions denote heavy-tailed consumption patterns.}
\label{fig:1a}
\end{figure}

\begin{figure}
\sidecaption
\includegraphics[width=.6\textwidth]{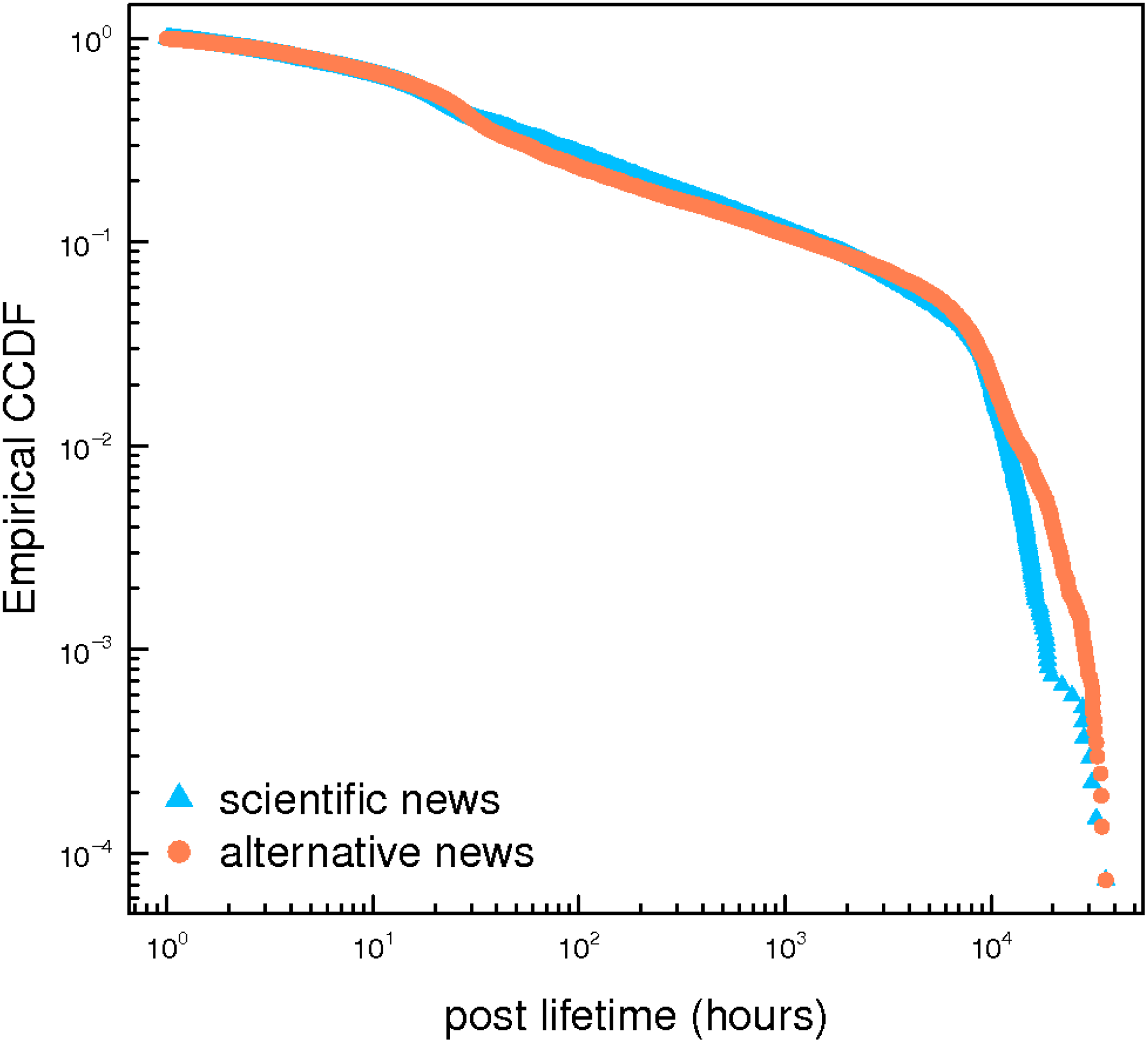}
\caption{\textsc{Italian Facebook}. Empirical CCDF, grouped by category, of the posts' lifetime i.e., the temporal distance (in hours) between the first and last comment. Lifetime is similar for both categories.}
\label{fig:1b}
\end{figure}

Fig.~\ref{fig:1a} shows the empirical Complementary Cumulative Distribution Functions (CCDFs) of users' activity on posts grouped by category on the Italian Facebook. We may notice that distributions of likes, comments, and shares are all heavy-tailed. To further investigate users' consumption patterns, in Fig.~\ref{fig:1b} we also plot the CCDF of the posts' lifetime, observing that distinct kinds of contents show a comparable lifetime.

\begin{figure}
\includegraphics[width=\textwidth]{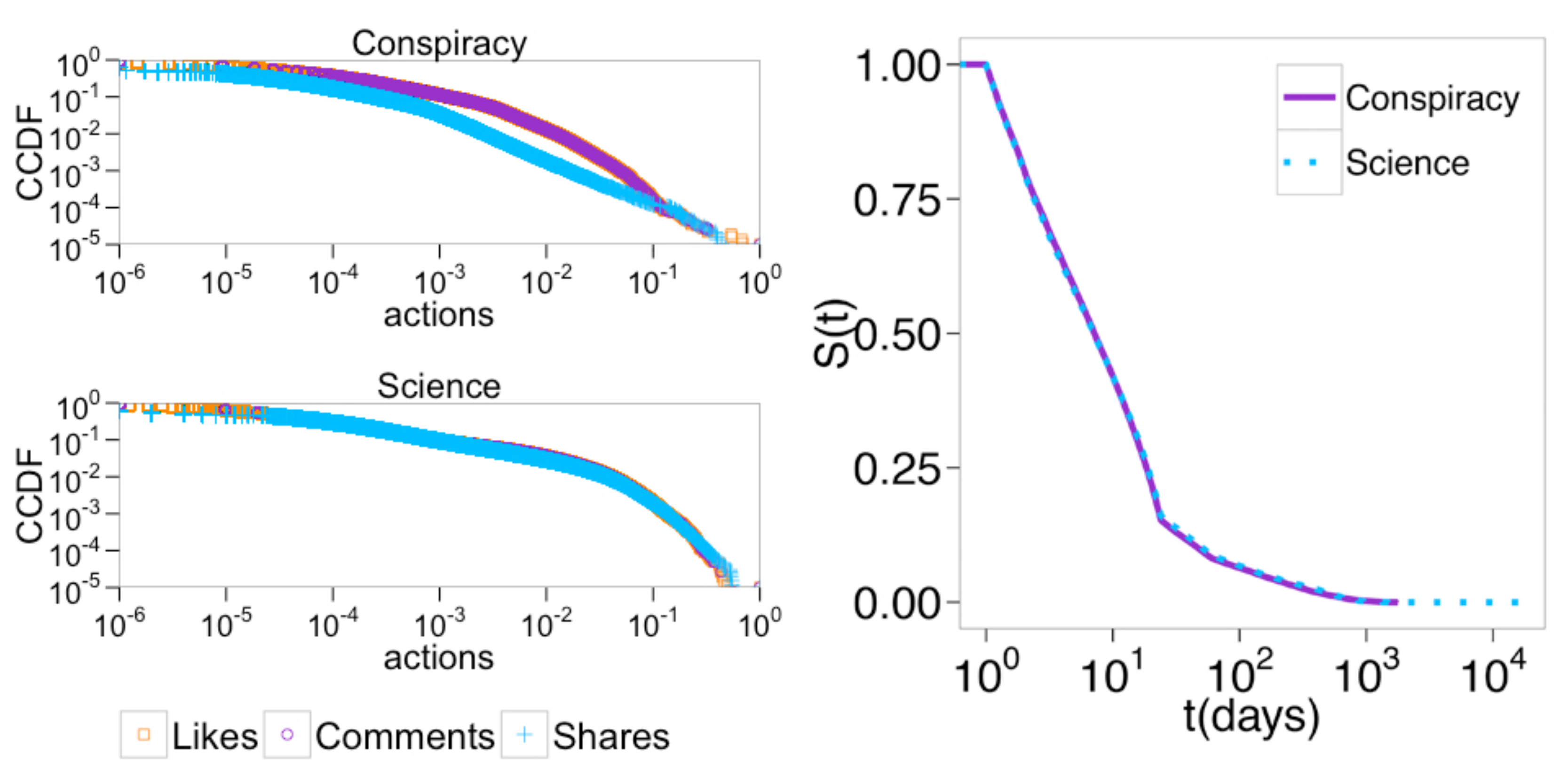}
\caption{\textsc{US Facebook} \textbf{Left}: Complementary cumulative distribution functions (CCDFs) of the number of likes, comments, and shares received by posts belonging to conspiracy \textit{(top)} and scientific \textit{(bottom)} news. \textbf{Right}: Kaplan-Meier estimates of survival functions of posts belonging to conspiracy and scientific news. Error bars are on the order of the size of the symbols.}
\label{fig:2}
\end{figure}

As for the US Facebook, the distribution of the number of likes, comments, and shares on posts belonging to both scientific and conspiracy news is shown in the left panel of Fig.~\ref{fig:2}. As seen from the plots, all distributions are heavy-tailed -- i.e, they are best fitted by power laws and possess similar scaling parameters. In the right panel of Fig.~\ref{fig:2}, we plot the Kaplan-Meier estimates of survival functions of posts grouped by category. To further characterize differences between the survival functions, we perform the Peto \& Peto \cite{Peto72asymptoticallyefficient} test to detect whether there is a statistically significant difference between the two survival functions. Since we obtain a p-value of $0.944$, we can state that there are not significant statistical differences between posts' survival functions on both science and conspiracy news. Thus, posts' persistence in the two categories is similar also in the US case.

Summarizing, our findings show that distinct kinds of information (science, conspiracy) are consumed in a comparable way. However, when considering the correlation between couples of actions, we find that users of conspiracy pages are more prone to both share and like a post, denoting a higher level of commitment \cite{bessi2015science}. Conspiracy users are more willing to contribute to a wide diffusion of their topics of interest, according to their belief that such information is intentionally neglected by mainstream media. 

\subsection{Polarization}
We now want to understand if users' engagement with a specific kind of content can become a good proxy to detect groups of users sharing the same system of beliefs i.e., echo chambers. Assume that a user $u$ has performed $x$ and $y$ likes (comments) on scientific and conspiracy posts, respectively, and let $\rho(u)=(y-x)/(y+x)$. Thus, we say that user $u$ is polarized towards science if $\rho(u)\le-0.95$, while she is towards conspiracy if $\rho(u)\ge0.95$ user $u$ is polarized towards conspiracy. 

\begin{figure}
\includegraphics[width=\textwidth]{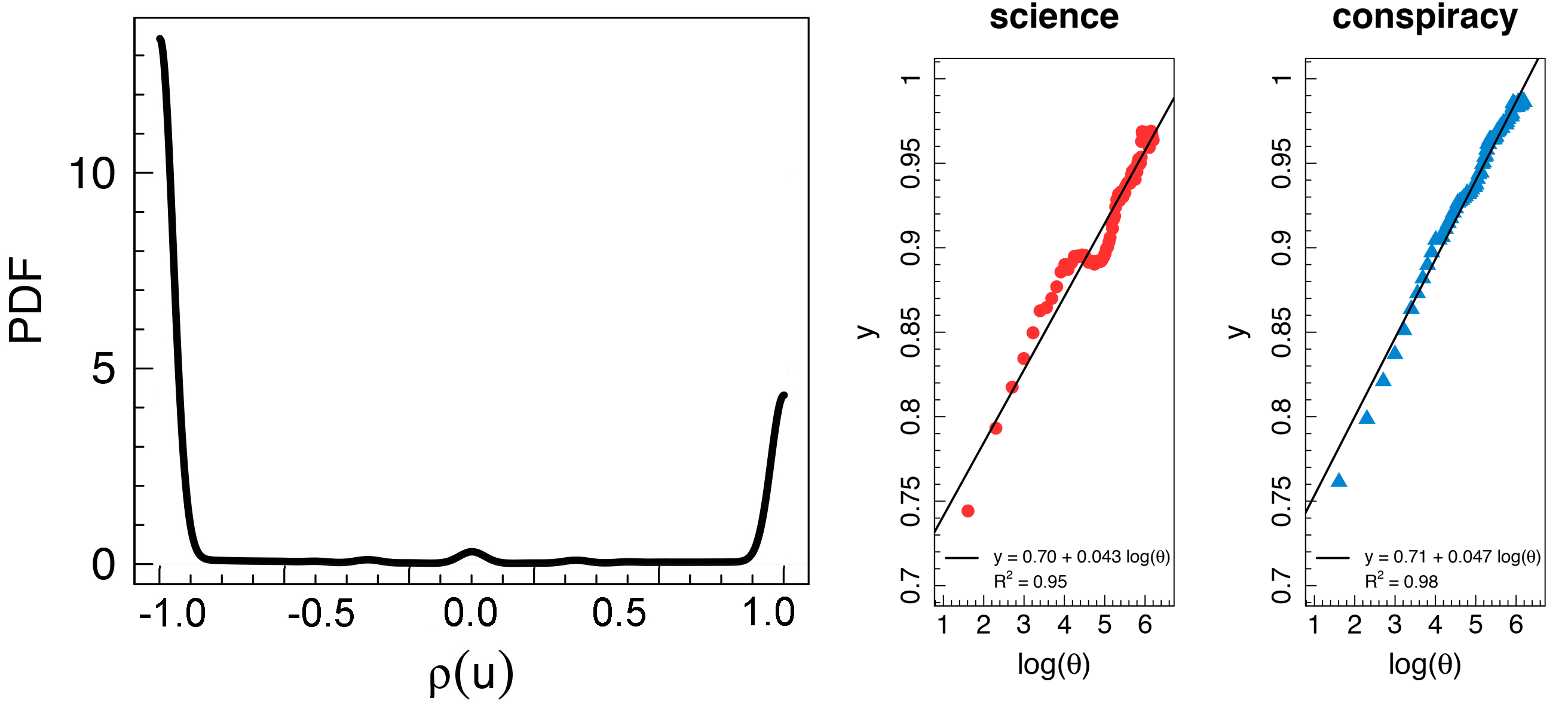}
\caption{\textsc{Italian Facebook} \textbf{Left}: Probability density function (PDF) of users' polarization. Notice the strong bimodality of the distribution, with two sharp peaks localized at $-1\lesssim \rho(u) \lesssim -0.95$ (conspiracy users) and at $0.95\lesssim \rho(u) \lesssim 1$ (science users). \textbf{Right}: Fraction of polarized neighbors as a function of the engagement $\theta$ for both science (left) and conspiracy (right) users. }
\label{fig:3}
\end{figure}

In Fig.~\ref{fig:3} we show the Probability Density Function (PDF) of users' polarization on the Italian Facebook. We observe a sharply peaked bimodal distribution where the vast majority of users is polarized either towards science ($\rho(u) \sim 1$) or conspiracy ($\rho(u) \sim -1$). Hence, most of likers can be divided into two groups of users, those \textit{polarized towards science} and those \textit{polarized towards conspiracy} news.  

Let us consider now the fraction of friends $y$ of a user $u$ sharing the same polarization of $u$. We define the \emph{engagement} $\theta(u)$ of a user $u$ as her liking activity normalized with respect to the total number of likes in our dataset. We find that the more a user is active on her narrative, the more she is surrounded by friends sharing the same attitude. Such a pattern is shown in the right panels of Fig.\ref{fig:3}. Hence, social interactions of Facebook users are driven by homophily: users not only tend to be very polarized, but they also tend to be linked to users with similar preferences. Indeed, in both right panels of Figure~\ref{fig:3} we can observe that for polarized users the fraction of friends with the same polarization is very high ($\gtrsim 0.75$) and grows with the engagement. 

\begin{figure}
\includegraphics[width=\textwidth]{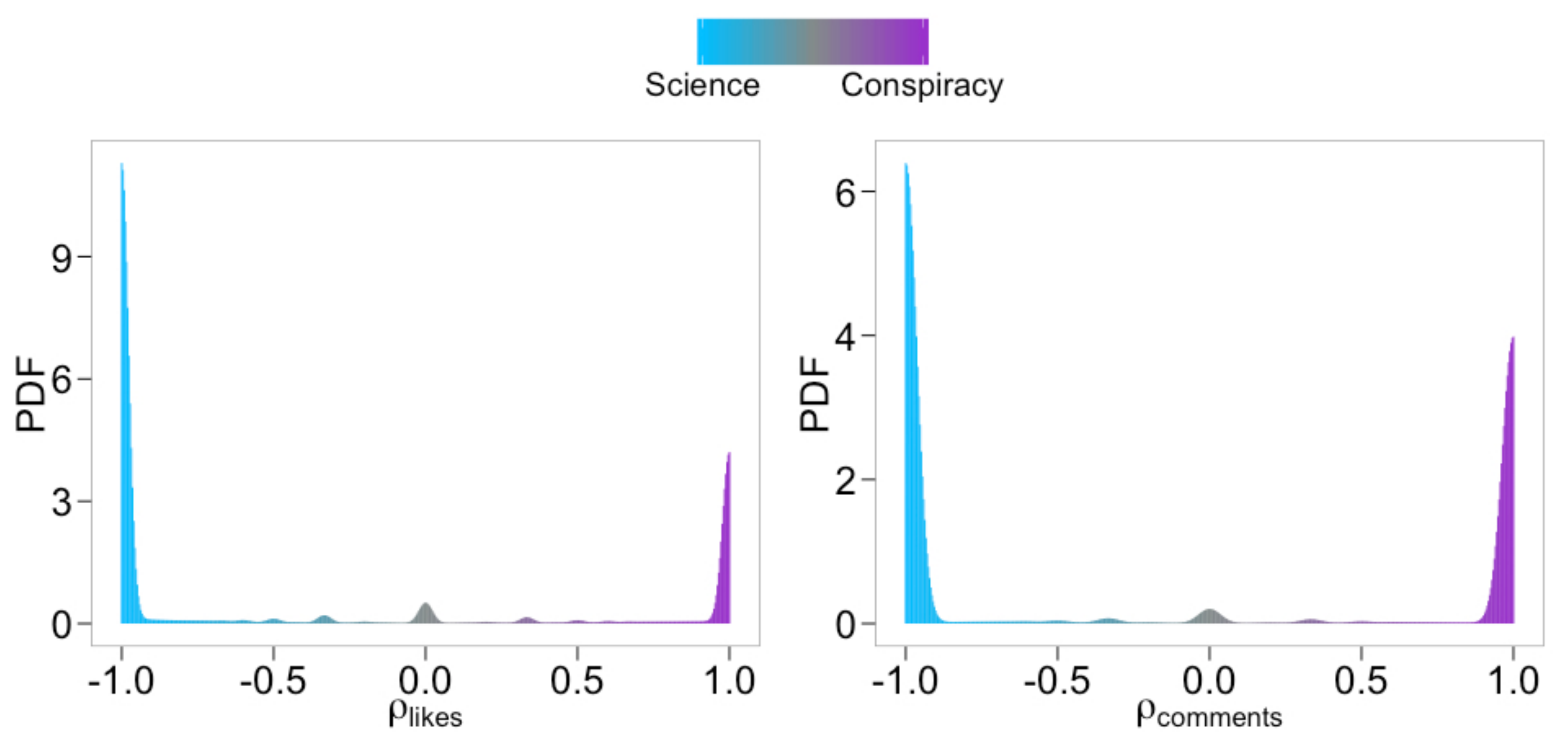}
\caption{\textsc{US Facebook} Probability Density Functions (PDFs) of the polarization of all users computed both on likes \textit{(left)} and comments \textit{(right)}.}
\label{fig:4}
\end{figure}

Similar patterns can be observed on the US Facebook. In Fig~\ref{fig:4} we show that the PDF for the polarization of all users is sharply bimodal here as well, with most having ($\rho(u) \sim -1$) or ($\rho(u) \sim 1$). Thus, most users may be divided into two main groups, those \emph{polarized towards science} and those \emph{polarized towards conspiracy}. The same pattern holds if we look at polarization based on comments rather than on likes.

In summary, our results confirm the existence of echo chambers on both the Italian and the US Facebook. Indeed, contents related to distinct narratives aggregate users into distinct, polarized communities, where users interact with like-minded people sharing their own system of beliefs. 

\section{Information Spreading and Cascades}
\label{sec:cascades}
\begin{figure}
\sidecaption
\includegraphics[width=.6\textwidth]{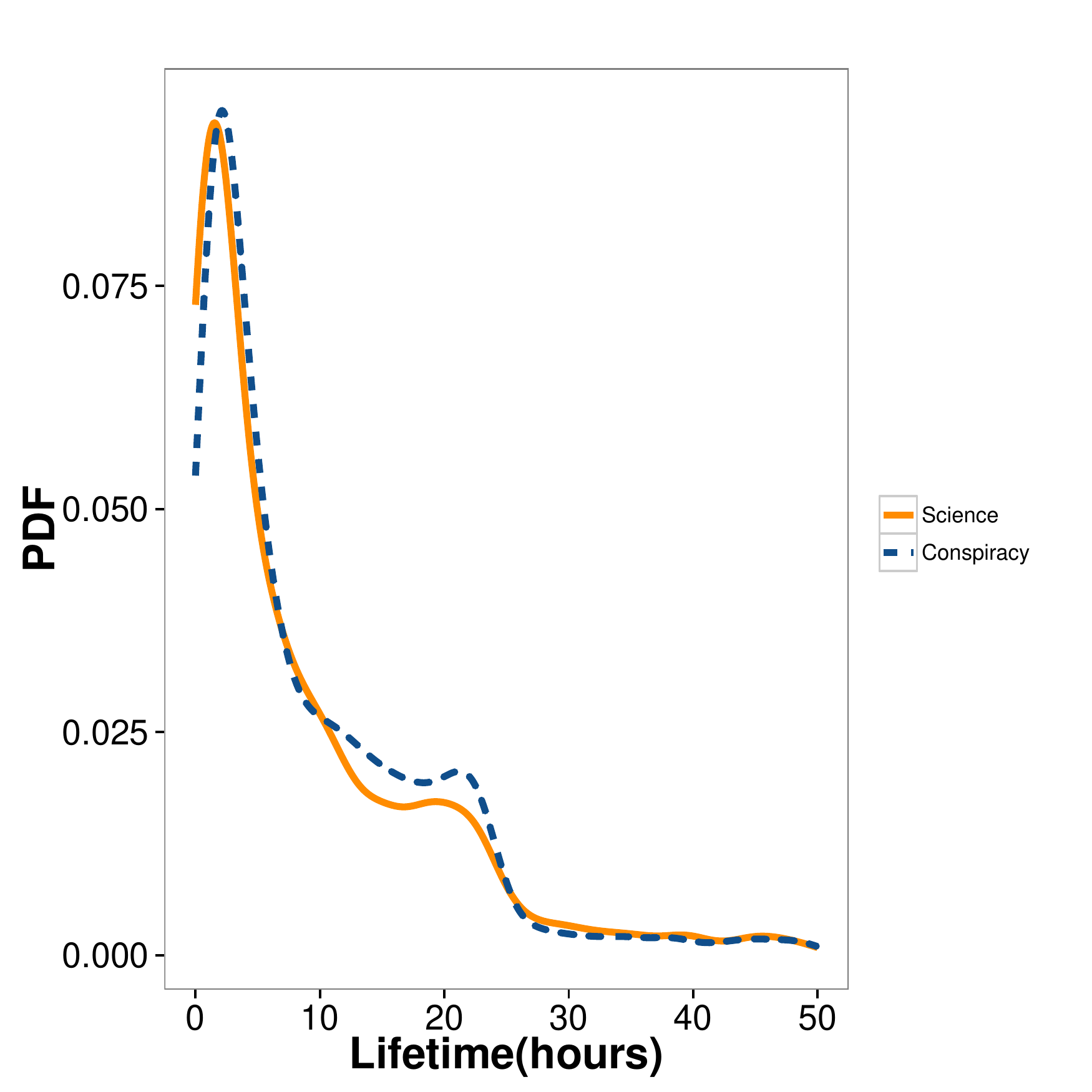}
\caption{\textsc{Italian Facebook} Probability Density Function
		(PDF) of lifetime computed on science news and conspiracy
		theories, where the lifetime is here computed as the
		temporal distance (in hours) between the first and
		last share of a post. Both categories show a similar
		behavior, with a peak in the first two hours and
		another around 20 hours.}
\label{fig:5}
\end{figure}

In this section we show how confirmation bias dominates viral processes of information diffusion and that the size of the (mis)information cascades may be approximated by the size of the echo chamber \cite{del2016spreading}. We begin our analysis by characterizing the statistical signature of cascades according to the narrative (science or conspiracy). Fig.~\ref{fig:5} shows the PDF of the cascade lifetime for both science and conspiracy. We compute the lifetime as the time (in hours) elapsed between the first and the last share of the post. In both categories we find a first peak at approximately 1--2 hours and a second peak at approximately 20 hours, denoting that the temporal sharing patterns are similar, independently of the narrative. We also find that a
significant percentage of the information spreads rapidly (24.42\% of
the science news and 20.76\% of the conspiracy rumors diffuse in less
than two hours, and 39.45\% of science news and 40.78\% of conspiracy
theories in less than five hours). Only 26.82\% of the diffusion of
science news and 17.79\% of conspiracy lasts more than one day.

\begin{figure}
\includegraphics[width=\textwidth]{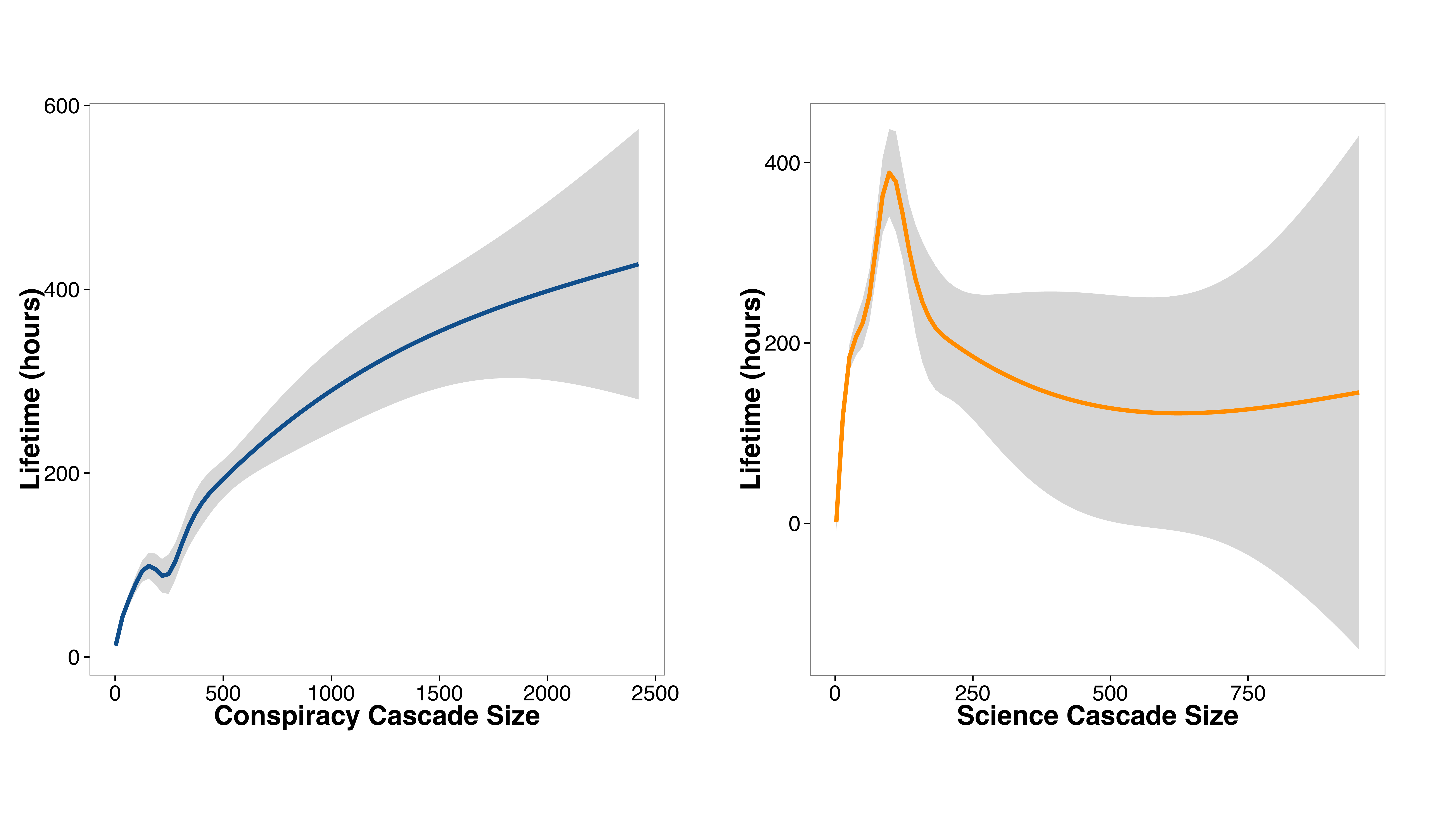}
\caption{\textsc{Italian Facebook} Lifetime as a function of the cascade's
		size for conspiracy news (left) and science news (right). We
		observe a contents-driven differentiation in the sharing
		patterns. For conspiracy the lifetime grows with the
		size, while for science news there is a peak in the lifetime
		around a value of the size equal to 200, and a higher
		variability in the lifetime for larger cascades.}
\label{fig:6}
\end{figure}

In Fig.~\ref{fig:6} we show the lifetime as a function of the cascade's size, i.e. the number of users sharing the post. For science news we observe a peak in the lifetime corresponding to a cascade's size value of $\approx 200$; moreover, the variability of the lifetime grows with the cascades' sizes, and higher cascade's size values correspond to high lifetime variability. For conspiracy-related contents, lifetime variability increases with cascade's size, and for highest values we observe a variability of the lifetime ~50\% around the average values. Such results suggest that news assimilation differs according to the categories. Science information is usually assimilated (i.e., it reaches a higher level of diffusion) quickly. A longer lifetime does not necessarily correspond to a higher level of interest, but possibly to a prolonged discussion within a specialised group of experts. Conversely, conspiracy rumors are assimilated more slowly and show a positive relation between lifetime and size; long-lived posts tend to be discussed by larger communities. 

\begin{figure}
\sidecaption
\includegraphics[width=.6\textwidth]{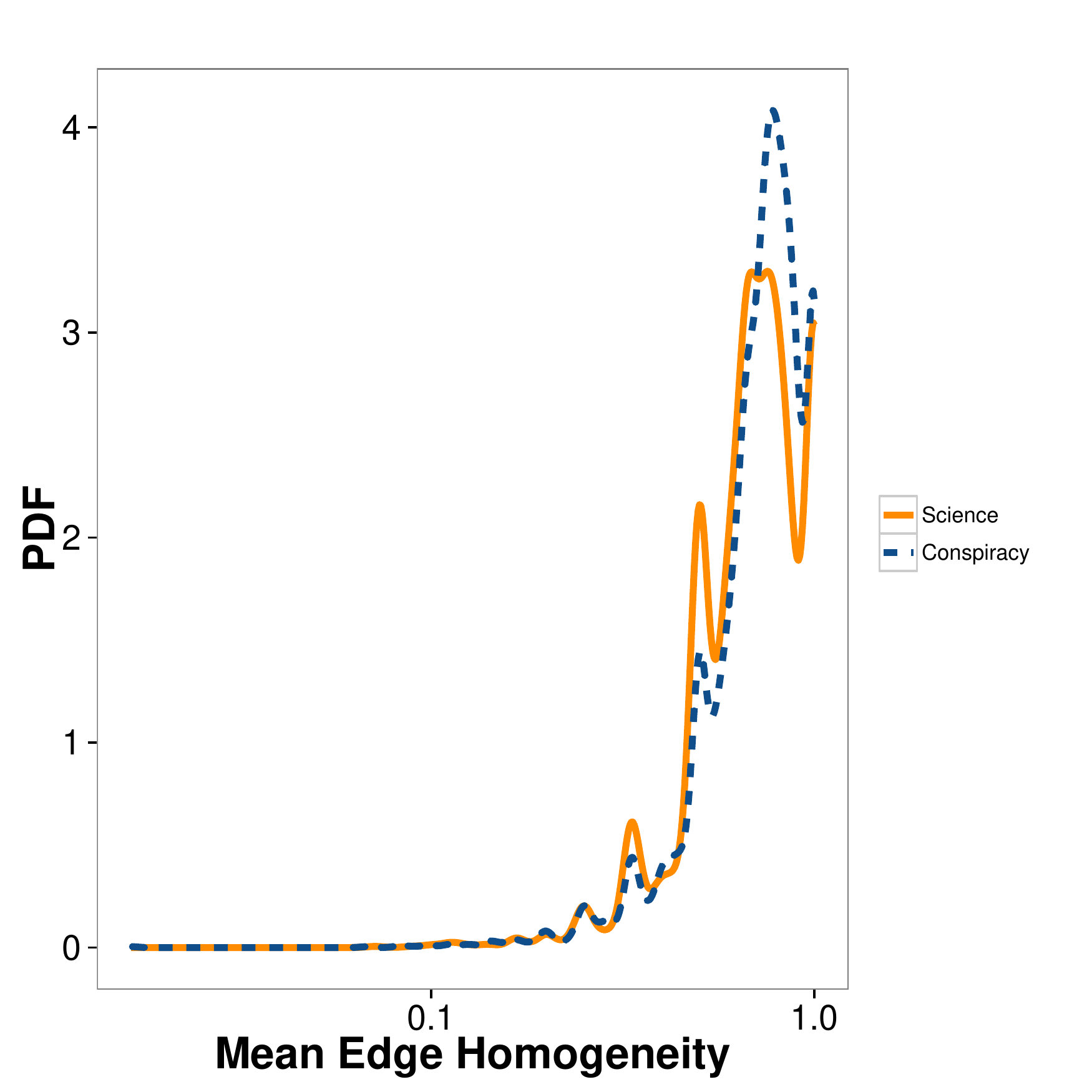}
\caption{\textsc{Italian Facebook} Mean edge homogeneity for science (solid orange) and conspiracy (dashed blue) news. The mean value of edge homogeneity on the whole sharing cascades is always greater or equal to zero.}
\label{fig:7}
\end{figure}

Finally, Fig.~\ref{fig:7} shows that the majority of links between consecutively sharing users is homogeneous, i.e. both users share the same polarization and, hence, belong to the same echo chamber. In particular, the average edge homogeneity value of all the observed sharing cascades is always greater than or equal to zero, suggesting that information spreading occurs mainly inside homogeneous clusters in which all users share the same polarization. Thus, contents tend to circulate only inside the echo chambers.

Summarizing, we found that cascades' dynamics differ, although consumption patterns on science and conspiracy pages are similar. Indeed, selective exposure is the primary driver of contents' diffusion and generates the formation of echo chambers, each with its own cascades' dynamics.

\section{Response to Paradoxical Information}
\label{sec:troll}
We have showed that users tend to aggregate around preferred contents shaping well-separated and polarized communities. Our hypothesis is that users' exposure to unsubstantiated claims may affect their selection criteria and increase their attitude to interact with false information. Thus, in this section we want to test how polarized users interact with information that is deliberately false i.e., troll posts, which are paradoxical imitations of conspiracy contents \cite{bessi2015science}. Such posts diffuse clearly dubious claims, such as the undisclosed news that infinite energy has been finally discovered, or that a new lamp made of actinides (e.g. plutonium and uranium) will finally solve the lack of energy with less impact on the environment, or that chemical analysis reveal that chem-trails contain \textit{sildenafil citratum} (sold as the brand name Viagra).

\begin{figure}
\sidecaption
\includegraphics[width=.6\textwidth]{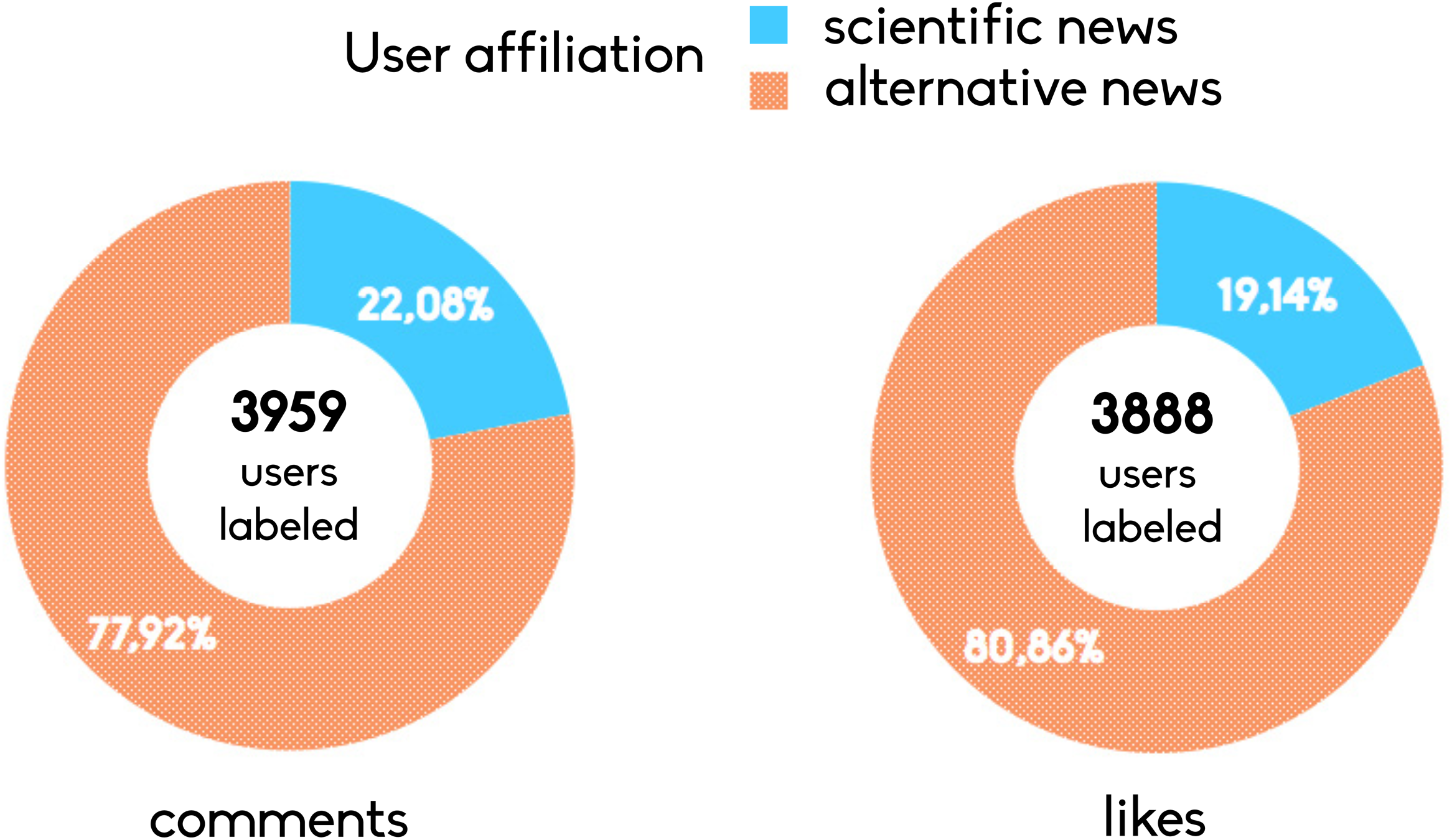}
\caption{\textsc{Italian Facebook} Percentage of comments and likes on troll posts from users polarized towards science (light blue) and conspiracy (orange).}
\label{fig:8}
\end{figure}

Fig.~\ref{fig:8} shows how polarized users of both categories interact with troll posts in terms of comments and likes on the Italian Facebook. Our findings show that users usually exposed to conspiracy claims are more likely to jump the credulity barrier: indeed, conspiracy users are more active in both liking and commenting troll posts. Thus, even when information is deliberately false and framed with a satirical purpose, its conformity with the conspiracy narrative transforms it into credible content for members of the conspiracy echo chamber. Evidently, confirmation bias plays a crucial role in content selection.

\section{Response to Dissenting Information}
\label{sec:debunking}
Debunking pages on Facebook strive to contrast misinformation spreading by providing fact-checked information to specific topics. However, if confirmation bias plays a pivotal role in selection criteria, then debunking is likely to sound to conspiracy users such as information dissenting from their preferred narrative. In this section, our aim is to study and analyze users' behavior w.r.t. debunking contents on the US Facebook \cite{zollo2015debunking}.
 
\begin{figure}
\sidecaption
\includegraphics[width=.6\textwidth]{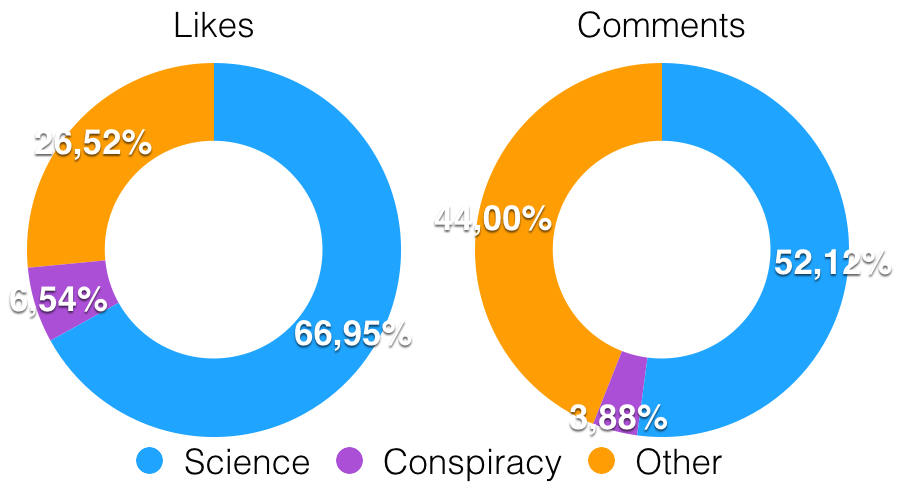}
\caption{\textsc{US Facebook} Proportions of likes \textit{(left)} and comments \textit{(right)} left by users polarized towards science, users polarized towards conspiracy, and not polarized users.}
\label{fig:9}
\end{figure}

As a first step, we show how debunking posts get liked and commented according to users' polarization. Fig.~\ref{fig:9} shows how users' activity is distributed on debunking posts: left (respectively, right) panel shows the proportions of likes (respectively, comments) left by users polarized towards science, users polarized towards conspiracy, and not polarized users. We notice that the majority of both likes and comments is left by users polarized towards science (respectively, $66,95\%$ and $52,12\%$), while only a small minority is made by users polarized towards conspiracy (respectively, $6,54\%$ and $3,88\%$). Indeed, the first interesting result is that the biggest consumer of debunking information is the scientific echo chamber. Out of $9,790,906$ polarized conspiracy users, just $117,736$ interacted with debunking posts -- i.e., commented a debunking post at least once.

\begin{figure}
\sidecaption
\includegraphics[width=.63\textwidth]{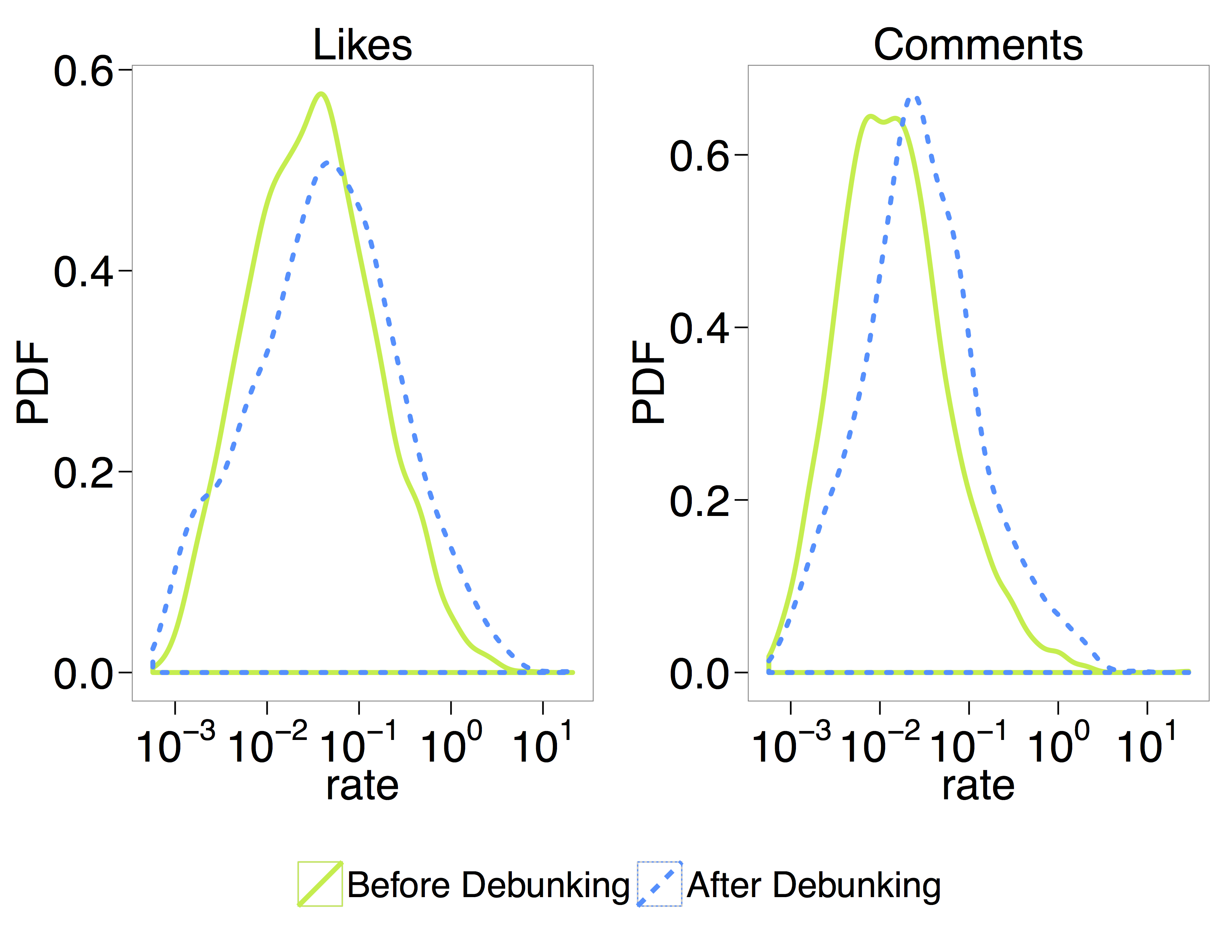}
\caption{\textsc{US Facebook} Rate -- i.e., average number, over time, of likes \textit{(left)} and comments \textit{(right)}) on conspiracy posts of users who interacted with debunking posts.}
\label{fig:10}
\end{figure}

Hence, debunking posts remain mainly confined within the scientific echo chamber and only few users usually exposed to unsubstantiated claims actively interact with the corrections. Dissenting information is mainly ignored. However, in our scenario few users belonging to the conspiracy echo chamber do interact with debunking information. We now wonder about the effect of such an interaction. Therefore, we perform a comparative analysis between users' behavior before and after they first comment on a debunking post. Fig.~ \ref{fig:10} shows the liking and commenting rates -- i.e, the average number of likes (or comments) on conspiracy posts per day -- before and after the first interaction with debunking. We can observe that users' liking and commenting rates increase after the interaction, Thus, their activity in the conspiracy echo chamber is reinforced. In practice, debunking attempts are acting as a backfire effect. 

\section{Emotional Dynamics}
\label{sec:emotional}
In this section, we aim at analyzing the emotional dynamics inside and across echo chambers. In particular, we apply sentiment analysis techniques to the comments of our Facebook Italian dataset, and study the aggregated sentiment with respect to scientific and conspiracy-like information \cite{zollo2015emotional}. The sentiment analysis is based on a supervised machine learning approach, where we first annotate a substantial sample of comments, and then build a Support Vector Machine (SVM) classification model. The model is then applied to associate each comment with one sentiment value: negative, neutral, or positive. The sentiment is intended to express the emotional attitude of Facebook users when posting comments.

\begin{figure}
\includegraphics[width=\textwidth]{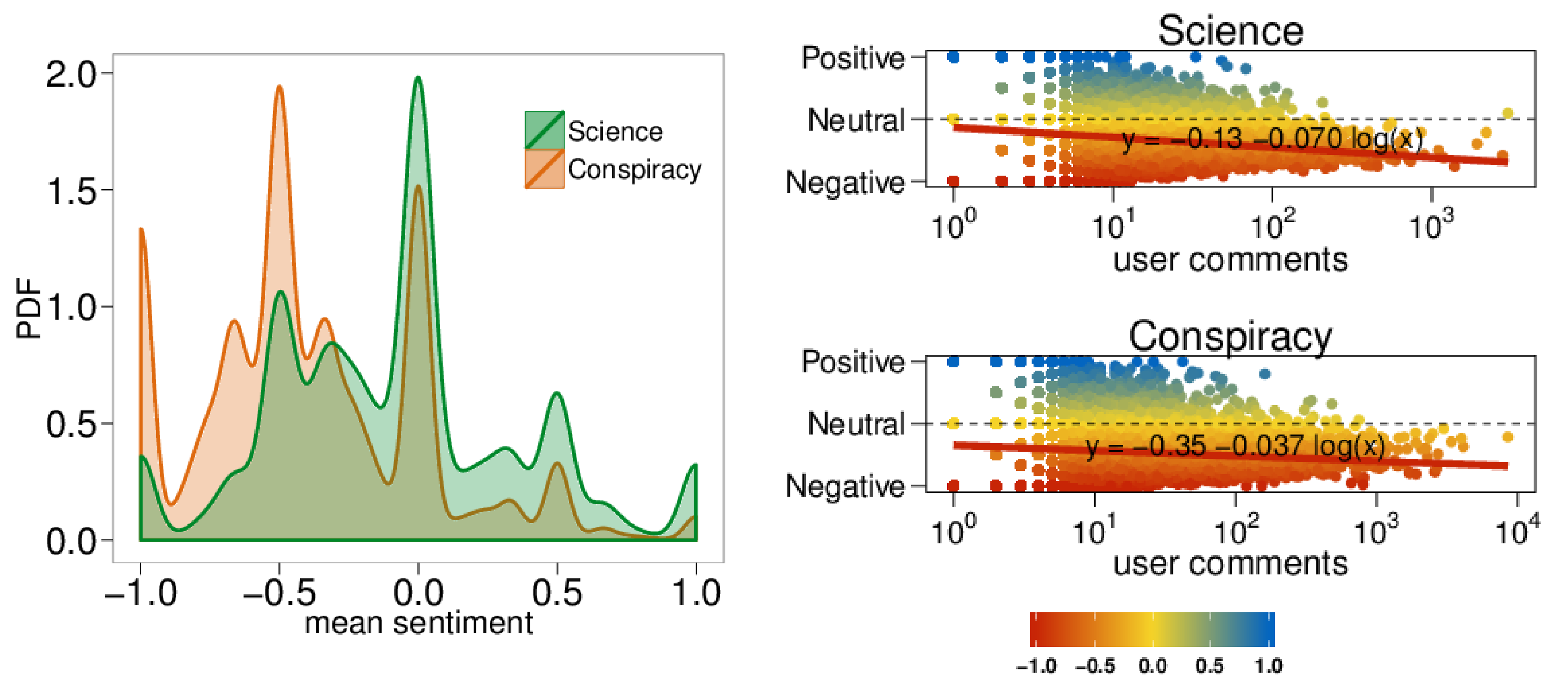}
\caption{\textsc{Italian Facebook} \textbf{Left}: Probability Density Function (PDF) of the mean sentiment of polarized users having commented at least twice, where $-1$ corresponds to negative sentiment, $0$ to neutral and $1$ to positive. \textbf{Right}:Average sentiment of polarized users as a function of their number of comments. Negative (respectively, neutral, positive) sentiment is denoted by red (respectively, yellow, blue) color. The sentiment has been regressed w.r.t. the logarithm of the number of comments.}
\label{fig:11}
\end{figure}

To further investigate the dynamics behind users' polarization, we now study how the sentiment changes w.r.t. users' engagement in their own echo chamber. 
In the left panel of Fig.~\ref{fig:11}, we show the PDF of the mean sentiment of polarized users with at least two comments. We may observe an overall negativity, more evident on the conspiracy side. When looking at the sentiment as a function of the number of comments of the user, we find that the more active a polarized user is, the more she tends towards negative values, both on science and conspiracy posts. Such results are shown in the right panel of Fig.~\ref{fig:11}, where the sentiment has been regressed w.r.t. the logarithm of the number of comments. Interestingly, the sentiment of science users decreases faster than that of conspiracy users.

\begin{figure}
\sidecaption
\includegraphics[width=.6\textwidth]{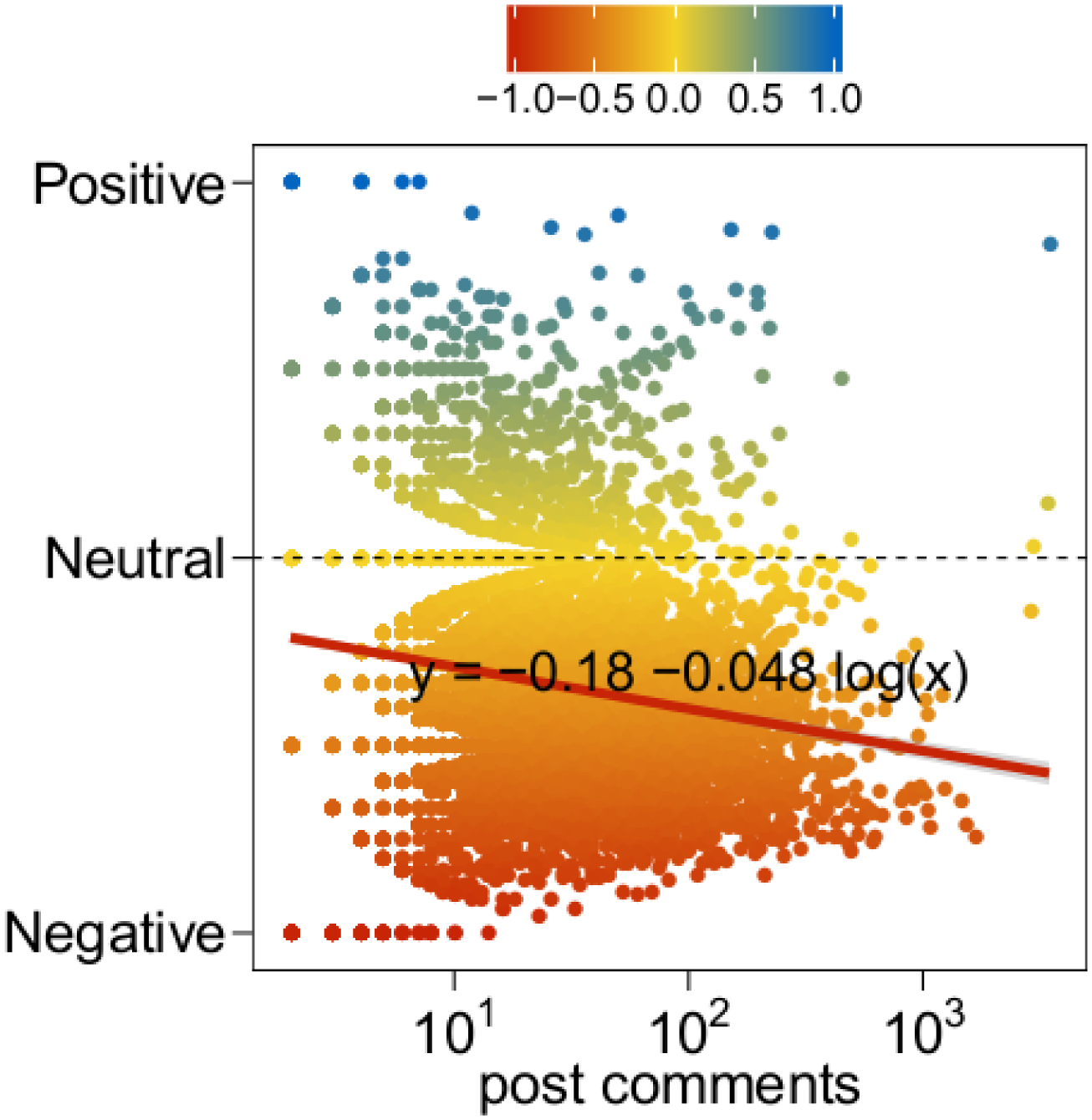}
\caption{\textsc{US Facebook} Aggregated sentiment of posts as a function of their number of comments. Negative (respectively, neutral, positive) sentiment is denoted by red (respectively, yellow, blue) color.}
\label{fig:12}
\end{figure}

We now want to investigate the emotional dynamics when such polarized (and negative-minded) users meet together. To this aim, we pick all the posts representing the arena where the debate between science and conspiracy users takes place. In particular, we select all the posts commented at least once by both a user polarized on science and a user polarized on conspiracy. We find $7,751$ such posts (out of $315,567$), reinforcing the fact that the two communities are strictly separated and do not often interact with one another. Then, we analyze how the sentiment changes when the number of comments of the post increases i.e., when the discussion becomes longer. Fig.~\ref{fig:12} shows the aggregated sentiment of such posts as a function of their number of comments. Clearly, as the number of comments increases -- i.e., the discussion becomes longer -- the sentiment is always more negative. Therefore, we may conclude that the length of the discussion does affect the negativity of the sentiment.

\section{Conclusions}
\label{sec:conclusions}
We investigated how information related to two very distinct narratives -- i.e., scientific and conspiracy news -- gets consumed and shapes communities on Facebook. For both the Italian and the US scenario, we showed the emergence of two well-separated and polarized groups -- i.e., echo chambers -- where users interact with like-minded people sharing the same system of beliefs. We found that users are extremely focused and self-contained on their specific narrative. Such a highly polarized structure facilitates the reinforcement and contents' selection by confirmation bias. Moreover, we observed that social interactions of Facebook users are driven by homophily: users not only tend to be very polarized, but they also tend to be linked to users with similar preferences. According to our results, confirmation bias dominates viral processes of information diffusion. Also, we found that the size of misinformation cascades may be approximated by the same size of the echo chamber.

Furthermore, by measuring the response to the injection of false information (parodistic imitations of alternative stories), we observed that users prominently interacting with alternative information sources -- i.e. more exposed to unsubstantiated claims -- are more prone to interact with intentional and parodistic false claims. Thus, our findings suggest that conspiracy users are more likely to jump the credulity barrier: even when information is deliberately false and framed with a satirical purpose, its conformity with the conspiracy narrative transforms it into credible content for members of the conspiracy echo chamber. 

Then, we investigated users' response to dissenting information. By analyzing the effectiveness of debunking on conspiracy users on the US Facebook, we found that scientific echo chamber is the biggest consumer of debunking posts. Indeed, only few users usually active in the conspiracy echo chamber interact with debunking information and, in the latter case, their activity in the conspiracy echo chamber increases after the interaction, rather than decreasing. Thus, debunking attempts are acting as a backfire effect.

Finally, we focused on the emotional dynamics inside and between the two echo chambers, finding that the sentiment of users on science and conspiracy pages tends to be negative, and is more and more negative when the discussion becomes longer or users' activity on the social network increase. In particular, the discussion degenerates when the two polarized communities interact with one another.  

Our findings provide insights about the determinants of polarization and the evolution of core narratives on online debating, suggesting that fact-checking is not working as expected. As long as there are no immediate solutions to functional illiteracy, information overload and confirmation bias will continue dominating social dynamics online. In such a context, misinformation risk and its consequences will remain significant. To contrast misinformation spreading, we need to smooth polarization.
To this aim, understanding how core narratives behind different echo chambers evolve is crucial and could allow to design more efficient communication strategies that account for users' cognitive determinants behind these kind of mechanisms.

\bibliographystyle{spmpsci}
\bibliography{biblio_springer.bib}

\begin{acknowledgement}
This work is based on co-authored material. We thank Aris Anagnostopoulos, Alessandro Bessi, Guido Caldarelli, Michela Del Vicario, Shlomo Havlin, Igor Mozeti\v{c}, Petra Kralj Novak, Fabio Petroni, Antonio Scala, Louis Shekhtman, H. Eugene Stanley, and Brian Uzzi.
\end{acknowledgement}

\end{document}